\documentclass[12pt,preprint]{aastex}
\usepackage{natbib}
\usepackage{lscape}
\begin{document}
\title{Stellar Velocities in the Carina, Fornax, Sculptor and Sextans dSph Galaxies: Data from the Magellan/MMFS Survey\footnote{This paper presents data gathered with the 6.5-m Magellan Telescopes at Las Campanas Observatory, Chile.}}
\shorttitle{MMFS Data}
\author{Matthew G. Walker\altaffilmark{1,2}, Mario Mateo\altaffilmark{2} and Edward W. Olszewski\altaffilmark{3}}
\email{walker@ast.cam.ac.uk}
\altaffiltext{1}{Institute of Astronomy, University of Cambridge, UK}
\altaffiltext{2}{Department of Astronomy, University of Michigan, Ann Arbor, MI}
\altaffiltext{3}{Steward Observatory, The University of Arizona, Tucson, AZ}

\begin{abstract} 

We present spectroscopic data for individual stars observed from 2004 March through 2008 August as part of our Michigan/MIKE Fiber System (MMFS) survey of four dwarf spheroidal (dSph) galaxies: Carina, Fornax, Sculptor and Sextans.  Using MMFS at the Magellan/Clay Telescope at Las Campanas Observatory, we have acquired 8855 spectra from 7103 red giant candidates in these Galactic satellites.  We list measurements of each star's line-of-sight velocity (median error $\pm 2.1$ km s$^{-1}$) and spectral line indices for iron and magnesium absorption features.  We use globular cluster spectra to calibrate the indices onto standard [Fe/H] metallicity scales, but comparison of the resulting metallicities with published values suggests that the MMFS indices are best used as indicators of relative, not absolute metallicity.  The empirical distributions of velocity and spectral indices also allow us to quantify the amount of contamination by foreground stars.  In a companion paper we develop an algorithm that evaluates the membership probability for each star, showing that the present MMFS sample contains more than $5000$ dSph members, including 774 Carina members, 2483 Fornax members, 1365 Sculptor members, and 441 Sextans members.  

\end{abstract}
\keywords{galaxies: dwarf ---  galaxies: kinematics and dynamics --- (galaxies:) Local Group ---  galaxies: individual (Carina, Fornax, Sculptor, Sextans) --- techniques: radial velocities}

\section{Introduction}

Dwarf spheroidal (dSph) galaxies are the smallest stellar systems thought to contain dark matter.  Because they represent the lower observational extreme of both the galaxy-luminosity and halo-mass functions, dSphs are important objects with which to compare models of galaxy formation.  As dSphs typically lack neutral hydrogen, their pressure-supported stellar components provide the best available kinematic tracers.  The dSph satellites of the Milky Way (MW) are sufficiently nearby that one can obtain high-resolution spectra of individual stars, enabling line-of-sight velocity measurements that resolve narrow (typically $\sim 5 -10$ km s$^{-1}$) dSph velocity dispersions.  The first such study used velocity measurements of just three stars to argue that the Draco dSph, if in virial equilibrium, has mass-to-light ratio $M/L \ge 30$ (solar units), indicative of a dominant dark matter component \citep{aaronson83}.  

Subsequent observations of Draco and other MW satellites supported the notion that dSph kinematics are dominated by dark matter.  Stellar velocity samples containing measurements for tens of stars per galaxy showed that the most luminous dSphs ($L_V\sim 10^{5-7}L_{V,\odot}$) all have central velocity dispersions of $\sim 10$ km s$^{-1}$ (e.g., \citealt{aaronson87,mateo91,mateo93,suntzeff93,hargreaves94a,armandroff95,hargreaves96b,edo95,queloz95,vogt95,mateo98b}).  Given such data, simple kinematic models that assume spherical symmetry, dynamic equilibrium, velocity isotropy and radially constant $M/L$ (i.e., mass follows light) imply that dSphs have masses of $\sim 10^7$ M$_{\odot}$ and $M/L \sim 10^{1-2}$.  The absence of a correlation between dynamical mass and luminosity suggests that the large variation in $M/L$ is solely a reflection of the variation in baryon content \citep{mateo93,mateo98b,gilmore07,walker07b,strigari08}.  



The advent of high-resolution, multi-object spectrographs at large telescopes now makes it possible to gather spectra for hundreds of stars during a single night.  With samples reaching hundreds of stars per dSph, velocity dispersion profiles are now available for all of the brighter MW dSphs \citep{kleyna02,kleyna04,wilkinson04,munoz05,munoz06a,sohn07,walker06a,walker06b,walker07b,battaglia06,mateo08,koch07a,koch07b}.  For the most luminous dSphs it is now possible to build spectroscopic data sets for thousands of stars.  With such large data sets one can measure higher moments (e.g., kurtosis) of the velocity distribution, and thereby place observational constraints on orbital anisotropy \citep{lokas05}.  The large data sets also provide information about the velocity distribution in two dimensions and thus are capable of uncovering kinematic evidence of substructure \citep{kleyna03,walker06b} as well as tidal streaming \citep{munoz06a,mateo08}.  The ability to measure spectral line strengths from these low S/N spectra provides an extra dimension of information that can be used to study stellar metallicity distributions (e.g., \citealt{koch06,koch07a}), to identify correlations between kinematics and metallicity \citep{tolstoy04,battaglia06}, and to help clean samples of contaminating foreground stars.  

In a previous paper \citep[ Paper I hereafter]{walker07a} we introduce a spectroscopic survey of individual dSph stars, undertaken using the Michigan/MIKE Fiber System (MMFS) at the Magellan 6.5-m Clay Telescope at Las Campanas Observatory.  Our MMFS spectra have resolution $R \sim 20000$ and sample the region $5140-5180$ \AA, which contains the prominent magnesium-triplet absorption feature.  Paper I describes MMFS as well as our procedures for target selection, observation and data reduction.  As of 2008 August we have used MMFS to obtain 8855 spectra from 7103 stars in four dSphs: Carina, Fornax, Sculptor and Sextans.  From each spectrum we measure the line-of-sight velocity and spectral indices that quantify the strengths of the iron and magnesium absorption lines present in the spectra (Paper I).  

Here we present the entire MMFS data set.  In a companion paper \citep[``Paper III'' hereafter]{walker08b} we develop a statistical algorithm that uses the available velocity and magnesium data, as well as the stellar positions, to evaluate for each star the probability of dSph membership.  Adding these probabilities, we find that the MMFS sample contains more than 5000 dSph members.  In forthcoming papers we use the MMFS data to provide detailed analyses of dSph kinematics and chemodynamic substructure.


\section{Observations \& Data}
\label{sec:data}

\begin{figure}
  \plotone{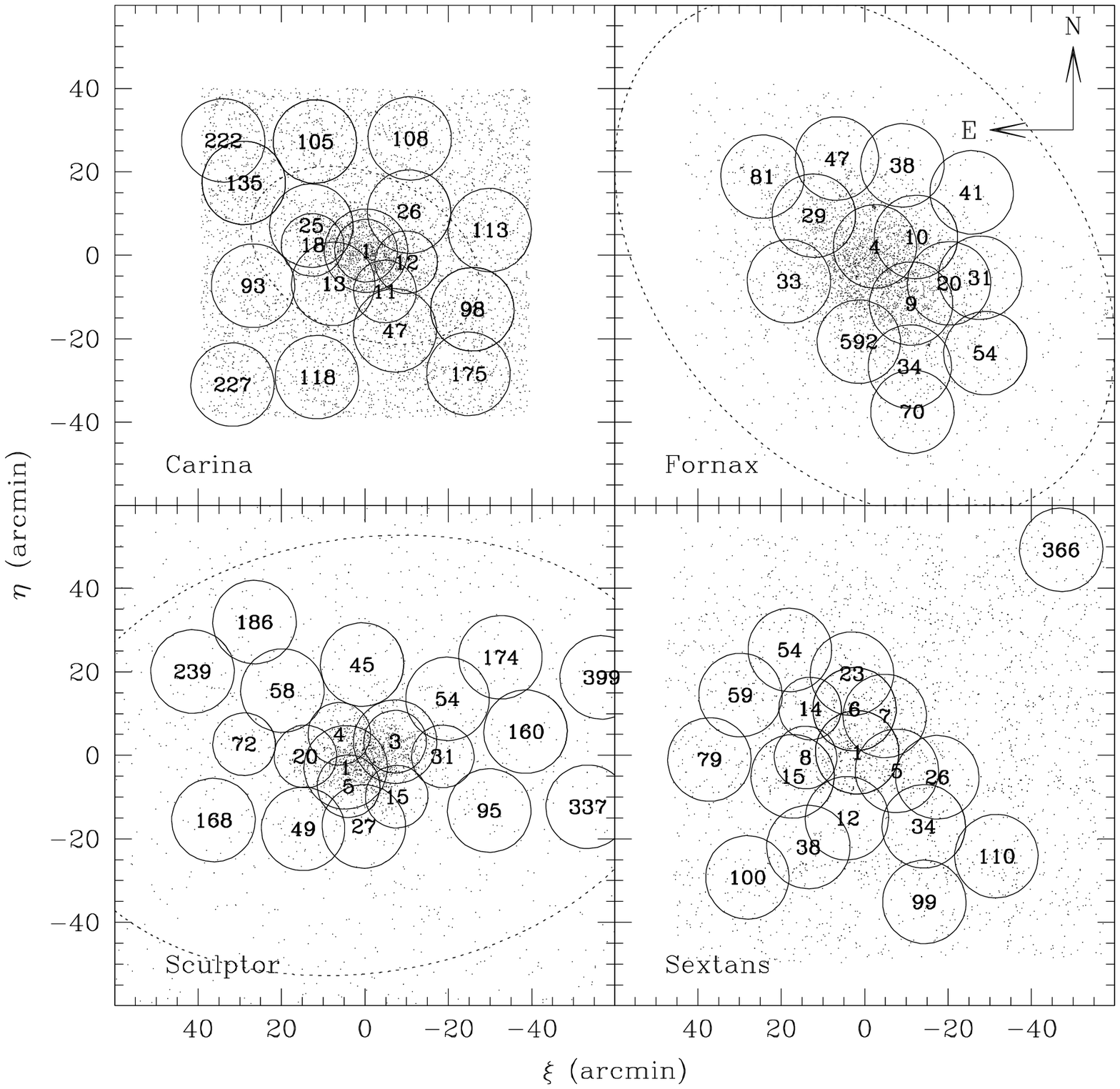}
  \caption{\scriptsize MMFS fields observed as of 2008 August.  Small points represent red giant candidates as identified in Paper I.  Dotted ellipses correspond to (nominal) tidal radii, as identified by \citet{king62} profile fits to the surface brightness profiles by \citet{ih95}.  Sextans' tidal radius ($r_t\sim 160\arcmin$) lies outside the plotted region.}
  \label{fig:newmaps}
\end{figure}

We refer the reader to Paper I for a description of MMFS and the details of our methodology regarding target selection, observing procedure and data reduction.  After the publication of Paper I we obtained new data during observing runs in 2007 January, 2007 September, 2008 April and 2008 August.  In these runs we used MMFS to observe nine additional Carina fields (including five distinct sets of targets in the densely populated central field), three Sextans fields, eight Sculptor fields and six Fornax fields.  Table \ref{tab:log} logs these new observations (see Paper I for a log of all previous MMFS observations).  The first two columns identify the galaxy and field number (identified in the maps of Figure \ref{fig:newmaps}).  Columns $3-7$ list the heliocentric Julian date at the midpoint of the first exposure, UT date at the midpoint of the first exposure, total exposure time, number of red giant candidates to which we assigned fibers, and the number of these for which we obtained an acceptable velocity measurement.  The fraction of observed stars with acceptable measurements suffered in both 2007 runs due to harsh observing conditions.  Figure \ref{fig:newmaps} maps all dSph fields observed with MMFS as of 2008 August.  

\renewcommand{\arraystretch}{0.6}
\begin{deluxetable}{lccccccl}
  \tabletypesize{\scriptsize}
  \tablewidth{0pc}
  \tablecaption{\scriptsize MMFS observations of dSph fields in 2007 - 2008}
  \tablehead{\colhead{Galaxy}&\colhead{Field}&\colhead{HJD $-2.45 \times 10^6$}&\colhead{UT Date}&\colhead{Exposure Time }&\colhead{Targets}&\colhead{Velocities}\\
    \colhead{}&\colhead{}&\colhead{(days)}&\colhead{}&\colhead{(s)}&\colhead{}&\colhead{}
  }
  \startdata
Carina&1.1&4120.7090&2007 Jan. 20&$4 \times 2400$&222&59\\
&&&&$1 \times 3600$&\nodata&\nodata\\
Carina&1.3&4122.7012&2007 Jan. 22&$3\times 2400$&210&121\\
Carina&1.4&4122.5869&2007 Jan. 22&$3\times 2400$&223&132\\
Sextans&1&4122.8218&2007 Jan. 22&$3\times 2000$&221&114\\
Carina&1.2&4123.5552&2007 Jan. 23&$3\times 2400$&218&70\\
Carina&98&4123.6812&2007 Jan. 23&$3\times 2400$&210&75\\
Sextans&59&4123.7988&2007 Jan. 23&$3\times 2400$&64&51\\
Carina&105&4124.6626&2007 Jan. 24&$4\times 2400$&174&70\\
Sextans&6&4124.8135&2007 Jan. 24&$3\times 2400$&196&109\\
Carina&113&4125.5474&2007 Jan. 25&$4\times 2400$&180&104\\
Carina&227&4125.6919&2007 Jan. 25&$4\times 2400$&124&90\\
\\
Sculptor&1.1&4354.7495&2007 Sep. 11&$3\times 2700$&224&59\\
Fornax&4.5&4355.7222&2007 Sep. 12&$3\times 3000$&224&128\\
Sculptor&3.1&4357.7109&2007 Sep. 14&$3 \times 2700$&218&55\\
Fornax&9.6&4359.6768&2007 Sep. 16&$3\times 1800$&222&96\\
Sculptor&3.2&4363.6250&2007 Sep. 20&$5\times 2700$&224&101\\
Sculptor&3.3&4364.6323&2007 Sep. 21&$2\times 2700$&218&30\\
&&&&$1\times 1900$&\nodata&\nodata\\
\\
Carina&1.2&4554.5176&2008 Mar. 29&$5\times 2400$&67&218\\
Carina&1.1&4557.5176&2008 Apr. 1 &$4\times 2400$&102&223\\
&&&&$1\times 1800$&\nodata&\nodata\\
Carina&1.5&4560.5107&2008 Apr. 3 &$4\times 2700$&117&219\\
\\
Sculptor&168&4683.7222&2008 Aug. 5&$3\times 2700$&25&41\\
Fornax&4.6&4683.8616&2008 Aug. 5&$3\times 2400$&165&222\\
Sculptor&174&4684.7157&2008 Aug. 6&$3\times 2700$&32&41\\
Fornax&47&4684.8451&2008 Aug. 6&$3\times 2700$&178&215\\
Sculptor&1.2&4685.7294&2008 Aug. 7&$3\times 2400$&153&205\\
Fornax&38&4685.8491&2008 Aug. 7&$3\times 2700$&121&143\\
Sculptor&399&4686.7057&2008 Aug. 8&$3\times 2700$&11&26\\
Fornax&41&4686.8369&2008 Aug. 8&$3\times 2400$&60&68\\
&&&&$1\times 2100$&\nodata&\nodata\\
  \enddata
  \label{tab:log}
\end{deluxetable}

Tables \ref{tab:car_table}, \ref{tab:for_table}, \ref{tab:scl_table} and \ref{tab:sex_table} present the MMFS spectroscopic data from each individual observation of targets in Carina, Fornax, Sculptor and Sextans, respectively.  For stars with multiple velocity measurements, results from repeat observations are listed directly beneath the first measurement.  Otherwise, results are listed chronologically by time of observation.  Column 1 identifies each target by galaxy and ID number.  Column 2 gives the field number and spectrograph channel (``B'' for blue, ``R'' for red) in which the star was observed.  Column 3 lists the Heliocentric Julian Date of the observation.  Columns $4-5$ list equatorial coordinates (J2000.0).  Columns $6-7$ give the apparent $V$ magnitude and $V-I$ color, respectively\footnote{We use our own photometric data for Carina and Fornax (see Paper I).  Sculptor photometry was provided by Matthew Coleman (see \citealt{coleman05a}).  Because of problems flat-fielding our Sextans images, for Sextans stars we list magnitudes and colors only if the stars overlap with the sample of \citet{lee03}, and we report the \citet{lee03} value.}.  Column 8 lists the measured velocity in the heliocentric rest frame (HRF).  Columns $9-10$ list composite spectral indices $\Sigma$Fe and $\Sigma$Mg, respectively.  

\begin{figure}
  \plotone{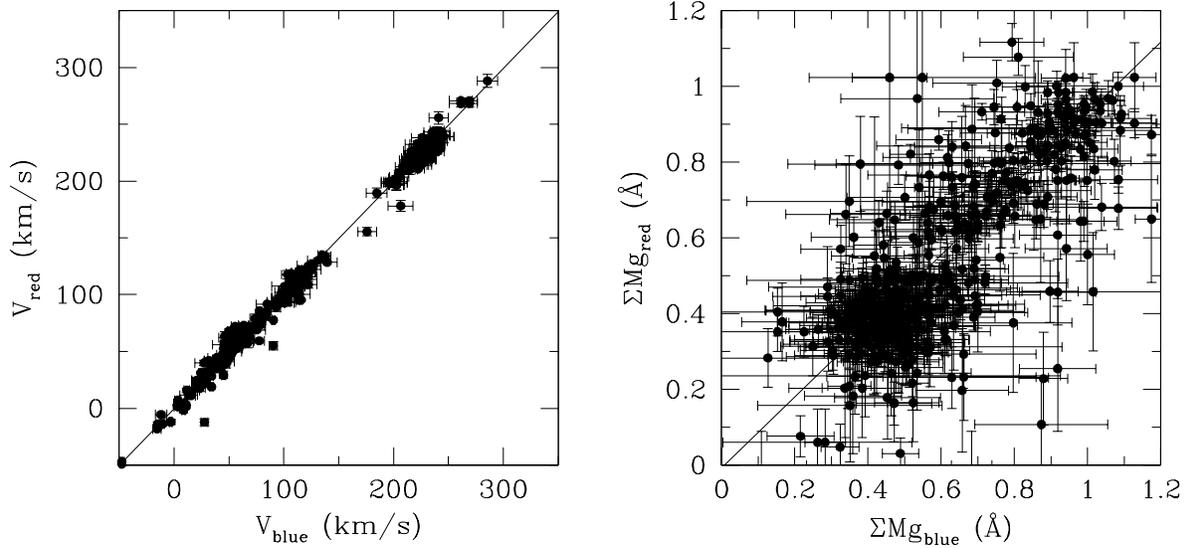}
  \caption[Comparison of blue- and red-channel measurements of composite Mg index]{\scriptsize \textit{Left:} Comparison of velocities measured for the same star on blue and red channels of the MIKE spectrograph.  \textit{Right:} Comparison of blue- and red-channel measurements of $\Sigma$Mg.  Overplotted in each panel is the best-fitting line.}  
  \label{fig:bluered}
\end{figure}

We use repeat measurements of stars measured on both blue and red channels of the MIKE spectrograph to quantify any systematic difference that may arise due to channel-dependent dispersion characteristics (\citealt{bernstein03}; Paper I).  For 647 stars we obtained at least one acceptable velocity measurement on both channels.  For these stars, the left-hand panel of Figure \ref{fig:bluered} plots the velocity measured on the blue channel against that measured on the red channel.  The best-fitting line through these data points has slope $0.998\pm 0.002$ and intercept $-0.33 \pm 0.52$ km s$^{-1}$, indicating there is no systematic difference between velocities measured with blue and red channels.  

We find evidence for a slight channel-dependence in measuring magnesium strength.  For 476 stars we obtained at least one acceptable measurement of $\Sigma$Mg on both channels.  For these stars, the right-hand panel in Figure \ref{fig:bluered} plots $\Sigma$Mg$_{blue}$ against $\Sigma$Mg$_{red}$.  The best-fitting straight line is given by 
\begin{equation}
  \Sigma\mathrm{Mg}_{blue}=(0.936 \pm 0.048)\Sigma \mathrm{Mg}_{red} - (0.007 \pm 0.048) \mathrm{\AA}.
  \label{eq:blueredindex}
\end{equation}
For subsequent analysis we apply Equation \ref{eq:blueredindex} to place values of $\Sigma$Mg measured with the red channel on the blue-channel scale.  The values of $\Sigma$Mg listed in Tables \ref{tab:car_table} - \ref{tab:sex_table} for red-channel spectra are those obtained \textit{after} applying Equation \ref{eq:blueredindex}.  We do not apply a similar correction, however, for values of $\Sigma$Fe measured with the red channel, because red-channel values of $\Sigma$Fe are calculated using fewer absorption lines than blue-channel values (the red channel has low throughput at the blue end of our spectral range; see Paper I).  Blue- and red-channel measurements of $\Sigma$Fe must therefore be considered separately.

\subsection{Metallicity Calibration}

\begin{figure}
  \plotone{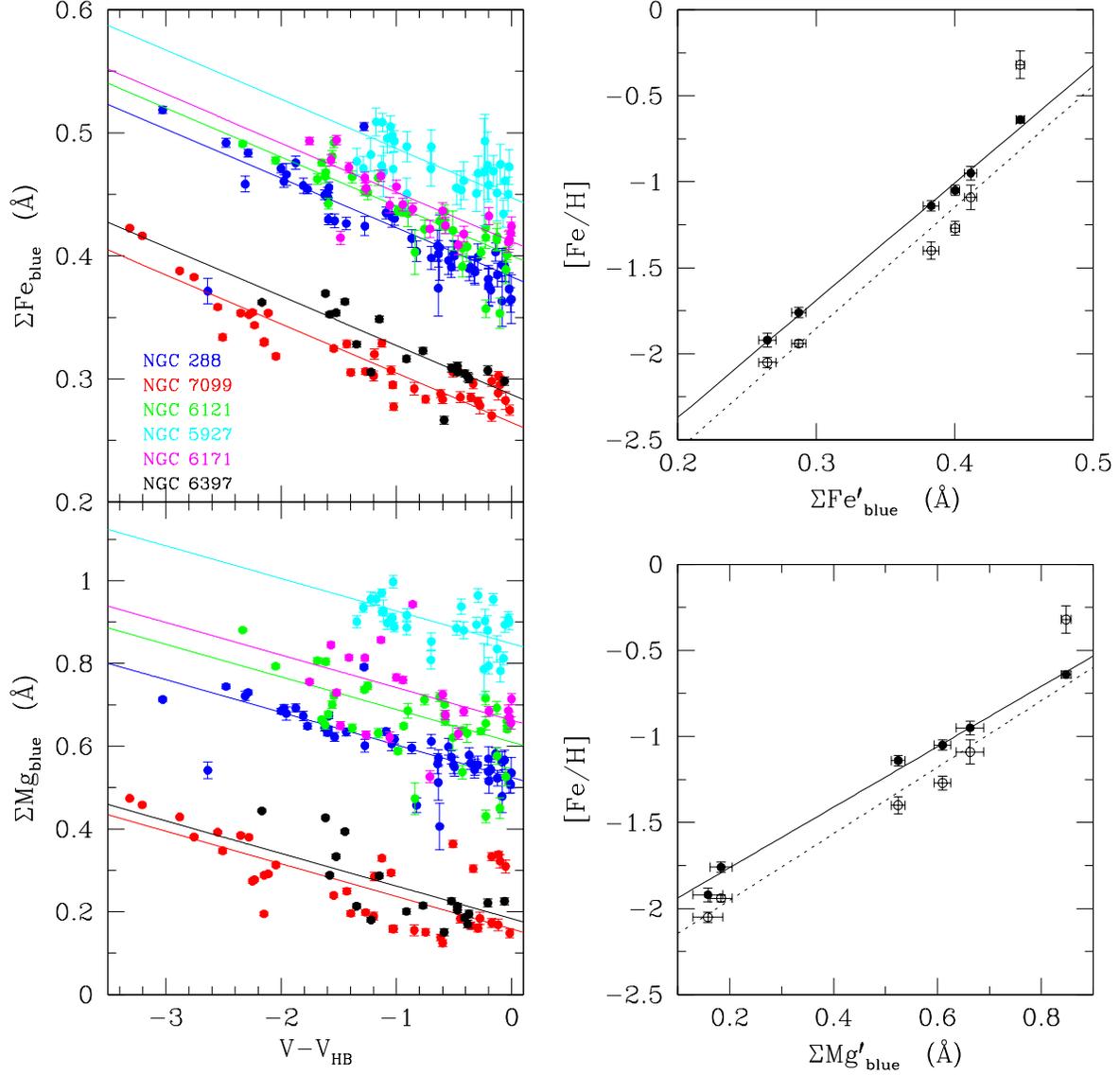}
  \caption{\scriptsize Metallicity calibration.  \textit{Left:} $\Sigma$Fe (top) and $\Sigma$Mg indices (bottom) versus $V-V_{HB}$ for globular cluster red giants observed with MMFS.  Overplotted are lines that result from fitting a single slope as well as a unique intercept for each cluster.  \textit{Right:} Mean reduced equivalent width versus [Fe/H] for the six clusters, on the scales of \citet{zinn84} (open circles) and \citet{carretta97} (solid circles).  Overplotted for each scale is the best-fitting straight line.}
  \label{fig:ind_calibrate}
\end{figure}

In principle the $\Sigma$Fe and $\Sigma$Mg indices contain information about metal abundances.  In order to calibrate these values onto a metallicity scale we obtained MMFS spectra of individual stars in globular clusters spanning the metallicity range $-2 \leq \mathrm{[Fe/H]} \leq -0.5$.  Left-hand panels in Figure \ref{fig:ind_calibrate} plot $\Sigma$Fe (top) and $\Sigma$Mg (bottom) against $V-V_{HB}$, where $V_{HB}$ is the apparent magnitude of the cluster's horizontal branch.  The empirical relationship is approximately linear for each cluster, as is the case for more conventional indices derived from the calcium triplet (``CaT''; e.g., \citealt{armandroff91}).  The slope reflects the dependence of opacity on surface gravity and temperature, both of which correlate with $V-V_{HB}$ as an evolved star ascends the red giant branch.  Following the procedure of \citet{rutledge97b} (see also \citealt{koch06}), we assume a common slope and fit straight lines to the data from each cluster.  The intercepts, known as ``reduced'' indices, are given by
\begin{equation}
  \Sigma\mathrm{Fe_{blue}}'=\Sigma\mathrm{Fe_{blue}}+(0.040\pm 0.001)(V-V_{HB});
  \label{eq:ind_calibratefe}
\end{equation}
\begin{equation}
  \Sigma\mathrm{Mg_{blue}}'=\Sigma\mathrm{Mg_{blue}}+(0.079\pm 0.002)(V-V_{HB}).
  \label{eq:ind_calibratemg}
\end{equation}
and provide a measure of metal abundance that, insofar as the linear model is valid, is independent of surface gravity and temperature.  

Right-hand panels in Figure \ref{fig:ind_calibrate} plot cluster metallicity, on the standard scales of \citet[``ZW84'']{zinn84} and \citet[``CG97'']{carretta97}, against the mean reduced indices for each globular cluster.  The best-fitting straight lines are given by
\begin{equation}
  \mathrm{[Fe/H]_{ZW84}}=(7.05\pm 2.10)\Sigma\mathrm{Fe} -3.97\pm 2.03 [2.19];
  \label{eq:ind_calibrate2}
\end{equation}
\begin{equation}
  \mathrm{[Fe/H]_{ZW84}}=(1.93\pm 0.16)\Sigma\mathrm{Mg}' -2.34\pm 0.15 [0.24];
  \label{eq:ind_calibrate3}
\end{equation}
\begin{equation}
  \mathrm{[Fe/H]_{CG97}}=(6.81\pm 1.87)\Sigma\mathrm{Fe}' -3.73\pm 0.48 [0.89];
  \label{eq:ind_calibrate4}
\end{equation}
\begin{equation}
  \mathrm{[Fe/H]_{CG97}}=(1.76\pm 0.16)\Sigma\mathrm{Mg}' -2.11\pm 0.10 [0.20];
  \label{eq:ind_calibrate5}
\end{equation}
where the errors include measurement errors as well as residuals from the fit, and the value in square brackets is the median metallicity error from dSph stars in the MMFS sample.  The metallicities obtained from the iron indices carry large errors due to the relative weakness of the iron features in the MMFS spectra.  The magnesium index thus provides our best measure of metallicity, yielding median errors of 0.20 dex on the scale of CG97 (0.24 dex on the scale of ZW84).  In subsequent discussion we consider only the [Fe/H] values on the CG97 scale, obtained using Equation \ref{eq:ind_calibrate5} (results are qualitatively unchanged if we use instead the values on the ZW84 scale, obtained from Equation \ref{eq:ind_calibrate3}).

\begin{figure}

  \plotone{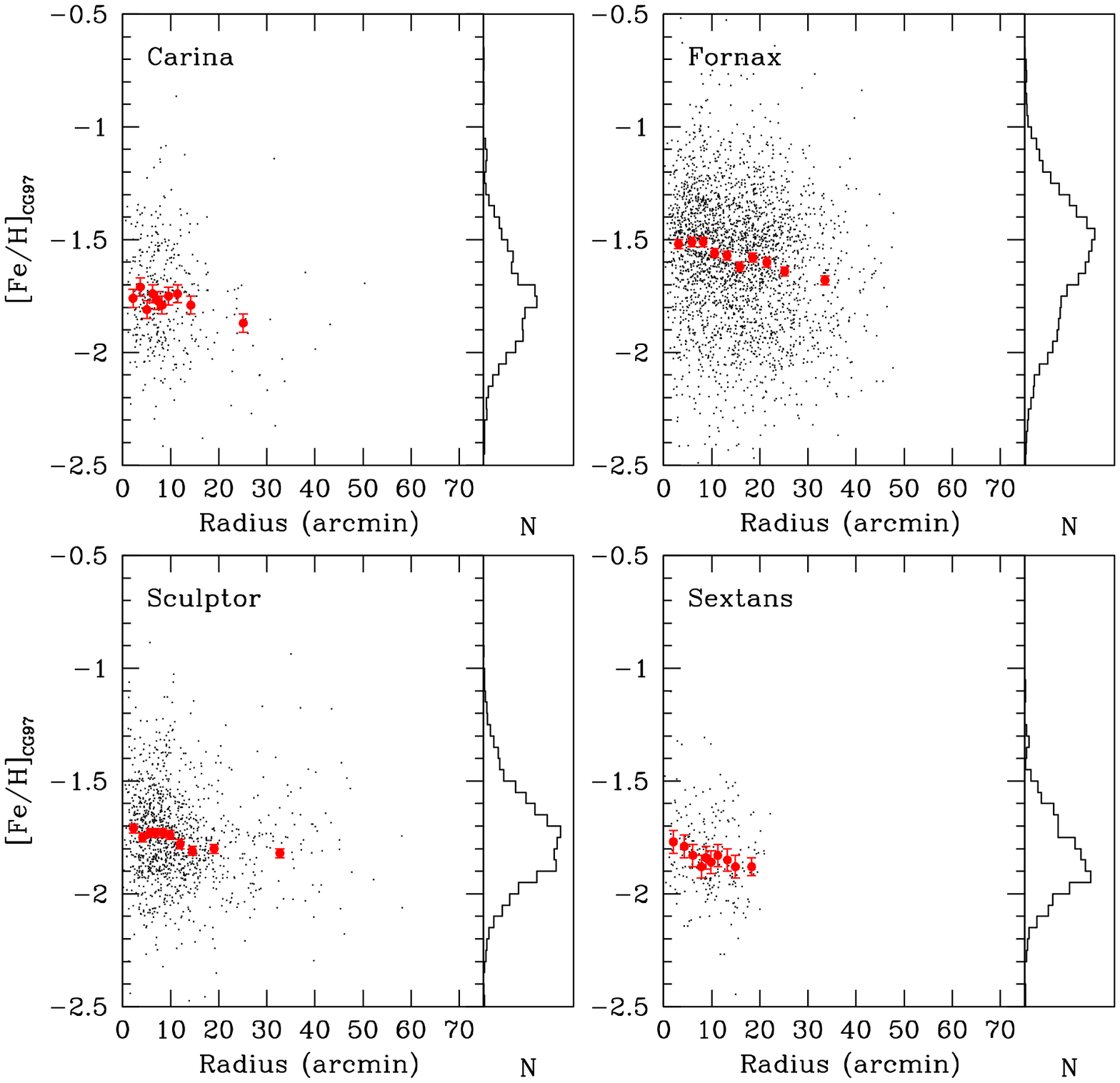}
  \caption{\scriptsize Metallicity distribution and radial dependence.  Black points in each panel give the [Fe/H] metallicity measured for individual stars, calibrated on the metallicity scale of \citet{carretta97}.  All plotted stars are red giant candidates with membership probability $\hat{P}_{dsph} >0.5$.  Median errors for individual data points are $\pm 0.2$ dex.  Red points give the mean [Fe/H] after binning the sample according to projected radius.  Histograms in subpanels give the global [Fe/H] distributions.}
  \label{fig:fehr}
\end{figure}

We apply Equations \ref{eq:ind_calibratefe} - \ref{eq:ind_calibrate5} to obtain [Fe/H] for each dSph star in the MMFS sample.  Figure \ref{fig:fehr} displays scatter plots of [Fe/H]$_{\mathrm{CG97}}$ against distance from the dSph center, as well as histograms of the global [Fe/H] distributions (results are qualitatively the same if we use the metallicity scale of ZW84).  We find that the resulting metallicity distributions are significantly narrower and biased toward high metallicity with respect to distributions previously derived from CaT spectroscopy.  For example, the Carina sample of \citet{koch06} and the Fornax sample of \citet{battaglia06} both contain significant numbers of stars with metallicities as low as [Fe/H]$\sim -2.5$ dex, and as high as $\sim -0.5$ dex; for both galaxies the MMFS distributions fall to zero short of these extremes.  While the MMFS sample ranks dSphs by mean [Fe/H] in the same sequence as previously published studies, it displays a narrower spread.  These comparisons lead us to conclude that either the slope we have determined in our metallicity calibration (Equations \ref{eq:ind_calibrate2} - \ref{eq:ind_calibrate5}) is simply too shallow, which we may remedy with observation of more calibrating clusters, or there are problems with one or both of the methods by which which Ca and Mg is used to measure iron abundance.  For example, \citet{kirby08} have recently shown that a spectral synthesis method that fits model spectra directly to iron lines implies wider metallicity distributions than are indicated by the standard CaT method.  

For analysis of the MMFS metallicity data we therefore recommend against using the absolute values of [Fe/H] that result from calibration using Equations \ref{eq:ind_calibrate2} - \ref{eq:ind_calibrate5}.  Instead we recommend using the reduced indices $\Sigma$Fe and $\Sigma$Mg, which are obtained by applying Equations \ref{eq:ind_calibratefe} - \ref{eq:ind_calibratemg} to the raw indices listed in Tables \ref{tab:car_table} - \ref{tab:sex_table}, as measures of \textit{relative} metallicity.  In this way the MMFS sample can be divided into metal-rich and metal-poor subcomponents, as has been done in previous analyses of these and other dSphs (e.g., \citealt{tolstoy04,battaglia06}), and can be used to search for evidence of metallicity gradients (Walker et al.\ in prep.)


\subsection{Repeat Measurements}

The MMFS data set contains repeat velocity measurements for 1363 dSph target stars.  There are 3115 independent measurements of these stars, including up to five measurements for some stars.  For subsequent analyses we replace the $n_j$ measurements of the $j^{th}$ star with the weighted mean, 
\begin{equation}
  \langle V\rangle_{j}=\frac{\sum_{i=1}^{n_j}(w_{ij}V_{ij})}{\sum_{i=1}^{n_j}w_{ij}},
  \label{eq:weightedmeanvelocity}
\end{equation} 
where the weights, $w_{ij}=\sigma_{V,ij}^{-2}$, are defined by the velocity errors $\sigma_{V_{ij}}$.  The weighted mean provides an unbiased estimate of the true velocity, and has variance $\sigma_{\bar{V},j}^2=(\sum_{i=1}^{n_j}w_{ij})^{-1}$.  We again use the weighted mean to combine repeat measurements (of which there are 2408 for 1089 stars) of the composite index $\Sigma$Mg.  For all stars with $n_j > 1$ independent velocity measurements, column 12 of Tables \ref{tab:car_table} - \ref{tab:sex_table} lists the weighted mean velocity.  Column 13 gives the weighted mean value of $\Sigma$Mg for stars with multiple measurements.  
  
\subsection{dSph Membership}
\label{subsec:membership}

We expect some fraction of the stars in each dSph sample to be contaminants---i.e., Milky Way stars with magnitudes and colors satisfying our target selection criteria.  These stars tend to be either disk dwarfs or halo giants, both of which have broader velocity distributions than dSphs.  Foreground dwarfs also tend to have higher metallicity and surface gravity, boosting their $\Sigma$Mg values.  Thus, while the dSph members cluster into relatively narrow velocity distributions, contaminants tend to have a broader velocity distribution and systematically larger $\Sigma$Mg.  We can therefore use the available velocity and $\Sigma$Mg data to help evaluate whether a given star is a dSph member or a foreground contaminant.  

In Paper III we introduce a statistical technique called expectation maximization (EM).  There we develop an EM algorithm that uses the velocity and $\Sigma$Mg data, as well as the positions of the observed stars, to evaluate the membership probability, $P_{M}$, of each star.  These probabilities are listed in Column 11 of Tables \ref{tab:car_table}-\ref{tab:sex_table}.  By adding membership probabilities, we find that there are more than 5000 probable members in the MMFS sample, including 774 Carina members, 2483 Fornax members, 1365 Sculptor members, and 441 Sextans members.

\section{Summary}

We have presented all spectroscopic observations of dSph stars taken with MMFS as of 2008 August.  The MMFS sample more than doubles the number of velocity measurements for dSph stars, dating from Aaronson's (1983) initial kinematic study to the most recent published data from surveys using Keck and the VLT \citep{munoz06a,koch06,koch07a,koch07b,battaglia06,battaglia08}.  In forthcoming work we use the data presented here, as well as the membership probabilities, to analyze the kinematics, chemodynamics and potential substructure in these dSphs.

Data tables that present the complete MMFS data sets are provided in the electronic version of this article and we welcome their use.  We thank the staff at Las Campanas Observatory for generous and expert support.  MGW and MM thank the Horace H. Rackham Graduate School at the University of Michigan for generous support, including funds for travel to Magellan.  MM acknowledges support from NSF grants AST-0206081  0507453, and 0808043.  EO acknowledges support from NSF Grants AST-0205790, 0505711, and 0807498.  MGW acknowledges support from the STFC-funded Galaxy Formation and Evolution programme at the Institute of Astronomy, University of Cambridge.


\begin{landscape}
\renewcommand{\arraystretch}{0.6}
 \begin{deluxetable}{lrcrrrrrrrrrrrrrrrrrrr}
 \tabletypesize{\scriptsize}
 \tablewidth{0pc}
 \tablecaption{MMFS Spectroscopic Data---Carina\tablenotemark{a}}
 \tablehead{\colhead{Target}&\colhead{Field}&\colhead{HJD}&
 \colhead{$\alpha_{2000}$}&\colhead{$\delta_{2000}$}&\colhead{V}&\colhead{V-I}&
 \colhead{$V_{helio}$}&\colhead{$\Sigma$Fe}&
 \colhead{$\Sigma$Mg}&\colhead{$\hat{P}_{M}$}&
 \colhead{$\langle V \rangle_{helio} $}&\colhead{$\langle \Sigma$Mg$\rangle$}\\
 \colhead{}&\colhead{}&\colhead{$-2.45\times 10^6$}&
 \colhead{(hh:mm:ss)}&\colhead{(dd:mm:ss)}&\colhead{(mag)}&\colhead{(mag)}&
 \colhead{(km s$^{-1}$)}&\colhead{(\AA)}&
 \colhead{(\AA)}&\colhead{}&\colhead{(km s$^{-1}$)}&\colhead{(\AA)}\\
 }
 \startdata
Car-0001&  18B&  3087.533&06:42:17.94&$-50$:53:58.4& 19.90    &  1.14    &$  218.8\pm   2.0$&$ 0.42\pm  0.05$&$ 0.48\pm  0.10$&1.000&$  219.0\pm   2.0$&$ 0.49\pm  0.10$\\
&   1R&  4554.518&&&&&$  223.0\pm  10.3$&$ 0.21\pm  0.28$&$ 0.53\pm  0.30$\\
Car-0002&  18B&  3087.533&06:42:18.60&$-50$:54:27.6& 18.87    &  1.14    &$  235.4\pm   0.6$&$ 0.38\pm  0.02$&$ 0.45\pm  0.05$&1.000&$  235.5\pm   0.6$&$ 0.44\pm  0.04$\\
&   1R&  3088.572&&&&&$  237.3\pm   5.8$&$ 0.43\pm  0.02$&$ 0.46\pm  0.09$\\
&  25R&  3411.554&&&&&$  236.0\pm   6.2$&$ 0.37\pm  0.07$&$ 0.24\pm  0.15$\\
&   1R&  4554.518&&&&&$  241.3\pm   4.0$&$ 0.43\pm  0.09$&$ 0.46\pm  0.16$\\
Car-0003&  18B&  3087.533&06:42:25.74&$-50$:52:12.6& 18.84    &  1.10    &$   39.6\pm   0.9$&$ 0.39\pm  0.04$&$ 1.12\pm  0.03$&0.000\\
Car-0004&  18B&  3087.533&06:42:27.17&$-50$:53:07.4& 19.53    &  0.99    &$   57.1\pm   1.9$&$ 0.43\pm  0.05$&$ 1.02\pm  0.06$&0.000\\
Car-0005&  18B&  3087.533&06:42:26.13&$-50$:53:29.1& 17.99    &  1.35    &$    4.6\pm   2.1$&$ 0.40\pm  0.02$&$ 0.95\pm  0.03$&0.000&$    6.0\pm   1.8$&$ 0.95\pm  0.03$\\
&   1R&  4122.587&&&&&$    8.8\pm   3.1$&\nodata&\nodata\\
Car-0006&  18B&  3087.533&06:42:27.95&$-50$:53:37.4& 20.41    &  0.92    &$  218.0\pm   3.4$&\nodata &$ 0.87\pm  0.22$&0.964\\
Car-0007&  18B&  3087.533&06:42:25.17&$-50$:54:01.8& 20.42    &  1.02    &$  216.2\pm   5.0$&$ 0.30\pm  0.23$&$ 0.57\pm  0.25$&0.997\\
Car-0008&  18B&  3087.533&06:42:21.59&$-51$:00:21.4& 20.47    &  0.84    &$  225.1\pm   5.7$&$ 0.42\pm  0.08$&$ 0.31\pm  0.20$&1.000&$  225.4\pm   3.1$&$ 0.24\pm  0.10$\\
&  13B&  3410.547&&&&&$  225.6\pm   3.8$&$ 0.38\pm  0.05$&$ 0.21\pm  0.12$\\
Car-0009&  18B&  3087.533&06:42:25.55&$-50$:59:35.8& 19.43    &  1.05    &$  232.1\pm   1.9$&$ 0.42\pm  0.04$&$ 0.41\pm  0.09$&1.000&$  232.1\pm   1.9$&$ 0.41\pm  0.09$\\
&   1R&  4123.555&&&&&$  232.9\pm   9.2$&\nodata&\nodata\\
Car-0010&  18B&  3087.533&06:42:24.18&$-50$:56:12.6& 18.48    &  1.20    &$  223.2\pm   0.9$&$ 0.40\pm  0.02$&$ 0.39\pm  0.04$&1.000&$  223.1\pm   0.8$&$ 0.39\pm  0.04$\\
&  25R&  3411.554&&&&&$  222.5\pm   1.6$&$ 0.34\pm  0.04$&$ 0.38\pm  0.10$\\
Car-0011&  18B&  3087.533&06:42:27.59&$-50$:55:54.0& 20.50    &  0.80    &$  214.5\pm   3.4$&$ 0.37\pm  0.07$&$ 0.35\pm  0.19$&1.000&$  217.0\pm   2.3$&$ 0.44\pm  0.09$\\
&  13B&  3410.547&&&&&$  219.1\pm   3.2$&$ 0.35\pm  0.06$&$ 0.47\pm  0.11$\\
Car-0012&  18B&  3087.533&06:42:11.82&$-50$:57:13.1& 19.94    &  1.00    &$  229.5\pm   4.0$&$ 0.33\pm  0.05$&$ 0.39\pm  0.12$&1.000\\
Car-0013&  18B&  3087.533&06:42:16.01&$-50$:56:54.8& 19.65    &  0.96    &$  219.1\pm   1.8$&$ 0.40\pm  0.03$&$ 0.35\pm  0.09$&1.000&$  219.3\pm   1.7$&$ 0.35\pm  0.09$\\
&   1R&  4122.587&&&&&$  226.7\pm  11.8$&\nodata&\nodata\\
Car-0014&  18B&  3087.533&06:42:14.01&$-50$:56:58.3& 20.62    &  0.95    &$  232.6\pm   2.5$&$ 0.47\pm  0.07$&$ 0.42\pm  0.16$&1.000\\
Car-0015&  18B&  3087.533&06:42:14.77&$-50$:56:21.5& 20.13    &  1.08    &$  220.3\pm   2.2$&$ 0.27\pm  0.07$&$ 0.52\pm  0.12$&0.999&$  220.5\pm   2.2$&$ 0.52\pm  0.12$\\
&   1R&  4557.518&&&&&$  224.8\pm  11.0$&\nodata&\nodata\\
...
 \enddata
 \tablenotetext{a}{See electronic edition for complete data table.}
 \label{tab:car_table}
 \end{deluxetable}

\renewcommand{\arraystretch}{0.6}
 \begin{deluxetable}{lrcrrrrrrrrrrrrrrrrrrr}
 \tabletypesize{\scriptsize}
 \tablewidth{0pc}
 \tablecaption{MMFS Spectroscopic Data---Fornax\tablenotemark{a}}
 \tablehead{\colhead{Target}&\colhead{Field}&\colhead{HJD}&
 \colhead{$\alpha_{2000}$}&\colhead{$\delta_{2000}$}&\colhead{V}&\colhead{V-I}&
 \colhead{$V_{helio}$}&\colhead{$\Sigma$Fe}&
 \colhead{$\Sigma$Mg}&\colhead{$\hat{P}_{M}$}&
 \colhead{$\langle V \rangle_{helio} $}&\colhead{$\langle \Sigma$Mg$\rangle$}\\
 \colhead{}&\colhead{}&\colhead{$-2.45\times 10^6$}&
 \colhead{(hh:mm:ss)}&\colhead{(dd:mm:ss)}&\colhead{(mag)}&\colhead{(mag)}&
 \colhead{(km s$^{-1}$)}&\colhead{(\AA)}&
 \colhead{(\AA)}&\colhead{}&\colhead{(km s$^{-1}$)}&\colhead{(\AA)}\\
 }
 \startdata
For-0001&  15B&  3287.826&02:39:53.25&$-34$:34:35.8& 19.27    &  1.35    &$   51.4\pm   0.4$&$ 0.51\pm  0.02$&$ 0.65\pm  0.05$&0.996\\
For-0002&  15B&  3287.826&02:39:56.10&$-34$:35:43.1& 19.23    &  1.27    &$   56.8\pm   0.5$&$ 0.51\pm  0.02$&$ 0.59\pm  0.05$&0.997\\
For-0003&  15B&  3287.826&02:39:55.28&$-34$:37:27.2& 19.06    &  1.24    &$   53.4\pm   0.4$&$ 0.50\pm  0.03$&$ 0.59\pm  0.07$&0.995\\
For-0004&  15B&  3287.826&02:39:55.69&$-34$:38:21.8& 19.22    &  1.28    &$   64.3\pm   0.6$&$ 0.56\pm  0.03$&$ 0.82\pm  0.05$&0.985\\
For-0005&  15B&  3287.826&02:39:58.47&$-34$:38:51.5& 18.97    &  1.17    &$   54.9\pm   0.5$&$ 0.49\pm  0.02$&$ 0.55\pm  0.05$&0.995\\
For-0006&  15B&  3287.826&02:39:55.36&$-34$:38:56.7& 19.33    &  1.13    &$   49.9\pm   0.8$&$ 0.49\pm  0.05$&$ 0.77\pm  0.08$&0.992\\
For-0007&  15B&  3287.826&02:39:59.50&$-34$:33:25.7& 19.24    &  1.25    &$   70.4\pm   0.5$&$ 0.57\pm  0.03$&$ 0.64\pm  0.06$&0.994\\
For-0008&  15B&  3287.826&02:40:01.52&$-34$:35:09.8& 19.30    &  1.32    &$   63.2\pm   0.5$&$ 0.62\pm  0.03$&$ 0.97\pm  0.05$&0.926\\
For-0009&  15B&  3287.826&02:40:00.13&$-34$:35:16.3& 19.38    &  1.19    &$   51.6\pm   0.5$&$ 0.54\pm  0.03$&$ 0.70\pm  0.07$&0.996\\
For-0010&  15B&  3287.826&02:39:58.41&$-34$:36:00.1& 19.03    &  1.28    &$   41.1\pm   0.4$&$ 0.53\pm  0.02$&$ 0.78\pm  0.04$&0.985\\
For-0011&  15B&  3287.826&02:40:00.18&$-34$:36:55.3& 19.18    &  1.17    &$   71.0\pm   1.6$&$ 0.49\pm  0.05$&$ 0.50\pm  0.15$&0.986\\
For-0012&  15B&  3287.826&02:40:07.09&$-34$:37:03.0& 19.24    &  1.08    &$   32.9\pm   1.2$&$ 0.35\pm  0.05$&$ 0.64\pm  0.10$&0.966\\
For-0013&  15B&  3287.826&02:40:07.35&$-34$:37:56.7& 19.06    &  1.26    &$   54.5\pm   0.6$&$ 0.58\pm  0.03$&$ 0.65\pm  0.09$&0.995\\
For-0014&  15B&  3287.826&02:40:05.29&$-34$:38:12.8& 19.14    &  1.31    &$   43.6\pm   1.0$&\nodata&\nodata &0.983\\
For-0015&  15B&  3287.826&02:40:11.40&$-34$:42:50.8& 19.22    &  1.13    &$   79.0\pm   0.8$&$ 0.43\pm  0.04$&$ 0.43\pm  0.09$&0.949&$   79.2\pm   0.8$&$ 0.42\pm  0.08$\\
& 592B&  3666.797&&&&&$   83.6\pm   3.2$&$ 0.38\pm  0.08$&$ 0.38\pm  0.19$\\
For-0016&  15B&  3287.826&02:40:14.65&$-34$:39:25.3& 19.31    &  1.21    &$   60.1\pm   0.8$&$ 0.47\pm  0.04$&$ 0.42\pm  0.09$&0.990\\
For-0017&  15B&  3287.826&02:40:12.08&$-34$:38:47.8& 19.25    &  1.34    &$   59.4\pm   0.5$&$ 0.45\pm  0.03$&$ 0.57\pm  0.06$&0.995\\
For-0018&  15B&  3287.826&02:40:23.40&$-34$:38:46.0& 19.18    &  1.23    &$   48.3\pm   0.7$&$ 0.43\pm  0.03$&$ 0.63\pm  0.06$&0.995\\
For-0019&  15B&  3287.826&02:40:25.26&$-34$:38:14.1& 19.20    &  1.04    &$  146.1\pm   5.7$&\nodata&\nodata &0.000\\
For-0020&  15B&  3287.826&02:40:04.61&$-34$:40:07.9& 19.32    &  1.36    &$   65.3\pm   0.6$&$ 0.50\pm  0.04$&$ 0.57\pm  0.08$&0.994\\
For-0021&  15B&  3287.826&02:39:56.33&$-34$:39:49.7& 19.11    &  1.30    &$   57.4\pm   0.7$&$ 0.50\pm  0.03$&$ 0.58\pm  0.07$&0.995\\
For-0022&  15B&  3287.826&02:40:01.03&$-34$:39:16.9& 18.92    &  1.13    &$   70.2\pm   0.7$&$ 0.44\pm  0.03$&$ 0.54\pm  0.06$&0.990\\
For-0023&  15B&  3287.826&02:40:09.10&$-34$:39:13.9& 19.24    &  1.16    &$   58.0\pm   0.6$&$ 0.50\pm  0.03$&$ 0.61\pm  0.07$&0.995\\
For-0024&  15B&  3287.826&02:40:04.67&$-34$:38:41.0& 19.22    &  1.15    &$   65.6\pm   0.6$&$ 0.54\pm  0.04$&$ 0.57\pm  0.08$&0.993\\
For-0025&  15B&  3287.826&02:40:02.44&$-34$:38:26.2& 19.17    &  1.18    &$   36.5\pm   0.6$&$ 0.51\pm  0.03$&$ 0.62\pm  0.07$&0.982\\
...
 \enddata
 \tablenotetext{a}{See electronic edition for complete data table}
 \label{tab:for_table}
 \end{deluxetable}

\renewcommand{\arraystretch}{0.6}
 \begin{deluxetable}{lrcrrrrrrrrrrrrrrrrrrr}
 \tabletypesize{\scriptsize}
 \tablewidth{0pc}
 \tablecaption{MMFS Spectroscopic Data---Sculptor\tablenotemark{a}}
 \tablehead{\colhead{Target}&\colhead{Field}&\colhead{HJD}&
 \colhead{$\alpha_{2000}$}&\colhead{$\delta_{2000}$}&\colhead{V}&\colhead{V-I}&
 \colhead{$V_{helio}$}&\colhead{$\Sigma$Fe}&
 \colhead{$\Sigma$Mg}&\colhead{$\hat{P}_{M}$}&
 \colhead{$\langle V \rangle_{helio} $}&\colhead{$\langle \Sigma$Mg$\rangle$}\\
 \colhead{}&\colhead{}&\colhead{$-2.45\times 10^6$}&
 \colhead{(hh:mm:ss)}&\colhead{(dd:mm:ss)}&\colhead{(mag)}&\colhead{(mag)}&
 \colhead{(km s$^{-1}$)}&\colhead{(\AA)}&
 \colhead{(\AA)}&\colhead{}&\colhead{(km s$^{-1}$)}&\colhead{(\AA)}\\
 }
 \startdata
Scl-0001&  20B&  3286.629&01:00:46.17&$-33$:39:19.6& 19.32    &  1.13    &$  114.1\pm   1.9$&$ 0.36\pm  0.02$&$ 0.43\pm  0.04$&0.998&$  112.6\pm   1.2$&$ 0.56\pm  0.03$\\
&   4B&  3288.571&&&&&$  111.5\pm   1.6$&$ 0.34\pm  0.02$&$ 0.69\pm  0.04$\\
Scl-0002&  20B&  3286.629&01:00:45.44&$-33$:39:59.9& 18.72    &  1.15    &$  110.1\pm   0.9$&$ 0.32\pm  0.01$&$ 0.39\pm  0.03$&1.000&$  110.2\pm   0.9$&$ 0.40\pm  0.03$\\
&   4R&  3288.571&&&&&$  111.5\pm   2.6$&$ 0.27\pm  0.03$&$ 0.42\pm  0.08$\\
Scl-0003&  20B&  3286.629&01:00:44.46&$-33$:41:54.1& 19.91    &  1.14    &$  118.4\pm   2.8$&$ 0.42\pm  0.04$&$ 0.53\pm  0.06$&1.000&$  116.4\pm   1.9$&$ 0.50\pm  0.05$\\
&   1B&  4017.555&&&&&$  114.8\pm   2.5$&$ 0.32\pm  0.05$&$ 0.39\pm  0.11$\\
Scl-0004&  20B&  3286.629&01:00:46.96&$-33$:43:19.9& 19.47    &  1.15    &$   95.6\pm   2.2$&$ 0.38\pm  0.02$&$ 0.34\pm  0.05$&1.000&$  103.8\pm   0.7$&$ 0.34\pm  0.03$\\
&   4B&  3288.571&&&&&$  104.5\pm   2.0$&$ 0.30\pm  0.02$&$ 0.32\pm  0.05$\\
&   1B&  4017.555&&&&&$  104.6\pm   0.8$&$ 0.36\pm  0.03$&$ 0.36\pm  0.07$\\
Scl-0005&  20B&  3286.629&01:00:42.44&$-33$:43:26.0& 19.11    &  1.18    &$  115.5\pm   2.1$&$ 0.43\pm  0.02$&$ 0.33\pm  0.04$&1.000&$  117.7\pm   0.9$&$ 0.40\pm  0.03$\\
&   4R&  3288.571&&&&&$  115.5\pm   7.9$&$ 0.39\pm  0.04$&$ 0.52\pm  0.10$\\
&   5R&  3289.593&&&&&$  114.6\pm   3.3$&$ 0.40\pm  0.03$&$ 0.53\pm  0.09$\\
&   1B&  4017.555&&&&&$  118.7\pm   1.1$&$ 0.38\pm  0.04$&$ 0.50\pm  0.08$\\
Scl-0006&  20B&  3286.629&01:00:44.88&$-33$:43:57.7& 19.17    &  1.11    &$   93.0\pm   2.3$&$ 0.38\pm  0.02$&$ 0.31\pm  0.05$&0.999&$   93.4\pm   2.2$&$ 0.32\pm  0.05$\\
&   4R&  3288.571&&&&&$  100.7\pm  10.1$&$ 0.15\pm  0.08$&$ 0.37\pm  0.14$\\
Scl-0007&  20B&  3286.629&01:00:42.07&$-33$:44:02.7& 19.26    &  1.18    &$  118.6\pm   2.0$&$ 0.29\pm  0.04$&$ 0.63\pm  0.05$&0.991\\
Scl-0008&  20B&  3286.629&01:00:51.01&$-33$:38:20.5& 18.97    &  1.13    &$  113.8\pm   4.9$&$ 0.30\pm  0.02$&$ 0.63\pm  0.03$&0.984&$  103.6\pm   1.4$&$ 0.60\pm  0.03$\\
&   1B&  4017.555&&&&&$  102.7\pm   1.5$&$ 0.33\pm  0.04$&$ 0.39\pm  0.08$\\
Scl-0009&  20B&  3286.629&01:00:50.99&$-33$:40:02.4& 18.50    &  1.17    &$   95.7\pm   0.8$&$ 0.39\pm  0.01$&$ 0.42\pm  0.03$&0.999&$   95.7\pm   0.8$&$ 0.41\pm  0.03$\\
&   4R&  3288.571&&&&&$   97.0\pm   4.1$&$ 0.33\pm  0.03$&$ 0.32\pm  0.09$\\
Scl-0010&  20B&  3286.629&01:00:49.48&$-33$:40:37.8& 18.98    &  1.16    &$  101.6\pm   1.6$&$ 0.33\pm  0.02$&$ 0.39\pm  0.03$&1.000&$  101.6\pm   1.6$&$ 0.41\pm  0.03$\\
&   4R&  3288.571&&&&&$  101.9\pm   5.1$&$ 0.40\pm  0.03$&$ 0.59\pm  0.09$\\
Scl-0011&  20B&  3286.629&01:00:51.15&$-33$:47:17.3& 19.39    &  1.04    &$  112.6\pm   4.8$&$ 0.34\pm  0.03$&$ 0.51\pm  0.05$&1.000&$  114.9\pm   1.1$&$ 0.49\pm  0.04$\\
&   1B&  4017.555&&&&&$  115.0\pm   1.1$&$ 0.37\pm  0.04$&$ 0.43\pm  0.09$\\
Scl-0012&  20B&  3286.629&01:00:51.95&$-33$:46:24.0& 19.12    &  1.20    &$  111.8\pm   0.8$&$ 0.36\pm  0.02$&$ 0.46\pm  0.04$&1.000&$  111.3\pm   0.4$&$ 0.45\pm  0.03$\\
&   5R&  3289.593&&&&&$  112.4\pm  11.2$&$ 0.30\pm  0.04$&$ 0.48\pm  0.10$\\
&   1B&  4017.555&&&&&$  111.0\pm   0.5$&$ 0.39\pm  0.03$&$ 0.39\pm  0.06$\\
...
 \enddata
 \tablenotetext{a}{See electronic edition for complete data table.}
 \label{tab:scl_table}
 \end{deluxetable}

\renewcommand{\arraystretch}{0.6}
 \begin{deluxetable}{lrcrrrrrrrrrrrrrrrrrrr}
 \tabletypesize{\scriptsize}
 \tablewidth{0pc}
 \tablecaption{MMFS Spectroscopic Data---Sextans\tablenotemark{a}}
 \tablehead{\colhead{Target}&\colhead{Field}&\colhead{HJD}&
 \colhead{$\alpha_{2000}$}&\colhead{$\delta_{2000}$}&\colhead{V}&\colhead{V-I}&
 \colhead{$V_{helio}$}&\colhead{$\Sigma$Fe}&
 \colhead{$\Sigma$Mg}&\colhead{$\hat{P}_{M}$}&
 \colhead{$\langle V \rangle_{helio} $}&\colhead{$\langle \Sigma$Mg$\rangle$}\\
 \colhead{}&\colhead{}&\colhead{$-2.45\times 10^6$}&
 \colhead{(hh:mm:ss)}&\colhead{(dd:mm:ss)}&\colhead{(mag)}&\colhead{(mag)}&
 \colhead{(km s$^{-1}$)}&\colhead{(\AA)}&
 \colhead{(\AA)}&\colhead{}&\colhead{(km s$^{-1}$)}&\colhead{(\AA)}\\
 }
 \startdata
Sex-0001&   8B&  3086.796&10:14:25.29&$-01$:42:53.7& 19.70    &  0.62    &$  269.4\pm   2.5$&$ 0.35\pm  0.03$&$ 0.45\pm  0.07$&0.000&$  270.2\pm   1.7$&$ 0.32\pm  0.06$\\
&  15B&  3417.761&&&&&$  270.9\pm   2.2$&$ 0.31\pm  0.03$&$ 0.14\pm  0.09$\\
Sex-0002&   8B&  3086.796&10:14:31.87&$-01$:39:06.6&\nodata   &\nodata   &$   47.3\pm   2.5$&$ 0.36\pm  0.02$&$ 0.99\pm  0.03$&0.000&$   47.1\pm   1.4$&$ 0.92\pm  0.02$\\
&  15B&  3417.761&&&&&$   47.1\pm   1.8$&$ 0.35\pm  0.02$&$ 0.84\pm  0.03$\\
Sex-0003&   8B&  3086.796&10:14:27.43&$-01$:37:24.8&\nodata   &\nodata   &$  123.8\pm   3.2$&$ 0.28\pm  0.03$&$ 1.05\pm  0.04$&0.000&$  123.3\pm   2.0$&$ 0.92\pm  0.03$\\
&  15B&  3417.761&&&&&$  122.9\pm   2.6$&$ 0.34\pm  0.02$&$ 0.80\pm  0.04$\\
Sex-0004&   8B&  3086.796&10:14:18.28&$-01$:44:08.4& 18.49    &  1.33    &$   29.8\pm   2.5$&$ 0.30\pm  0.05$&$ 1.00\pm  0.06$&0.000&$   30.1\pm   1.8$&$ 0.85\pm  0.03$\\
&  15B&  3417.761&&&&&$   30.4\pm   2.5$&$ 0.33\pm  0.02$&$ 0.80\pm  0.04$\\
Sex-0005&   8B&  3086.796&10:14:18.50&$-01$:39:59.3& 18.06    &  1.25    &$   31.9\pm   1.9$&$ 0.38\pm  0.02$&$ 0.98\pm  0.02$&0.000&$   32.4\pm   1.3$&$ 0.94\pm  0.01$\\
&  15B&  3417.761&&&&&$   32.8\pm   1.8$&$ 0.37\pm  0.01$&$ 0.89\pm  0.02$\\
Sex-0006&   8B&  3086.796&10:14:20.25&$-01$:39:23.0& 19.77    &  0.73    &$  273.5\pm   4.0$&$ 0.28\pm  0.05$&$ 0.48\pm  0.12$&0.000&$  271.6\pm   1.8$&$ 0.62\pm  0.05$\\
&  15B&  3417.761&&&&&$  271.1\pm   2.1$&$ 0.35\pm  0.03$&$ 0.64\pm  0.05$\\
Sex-0007&   8B&  3086.796&10:14:17.73&$-01$:38:28.7& 19.42    &  0.58    &$  147.6\pm   3.0$&$ 0.24\pm  0.04$&$ 0.48\pm  0.08$&0.000&$  145.9\pm   2.0$&$ 0.37\pm  0.05$\\
&  15B&  3417.761&&&&&$  144.5\pm   2.8$&$ 0.25\pm  0.03$&$ 0.30\pm  0.06$\\
Sex-0008&   8B&  3086.796&10:14:06.42&$-01$:34:50.0& 17.55    &  1.26    &$   -9.0\pm   1.7$&$ 0.35\pm  0.01$&$ 0.87\pm  0.02$&0.000\\
Sex-0009&   8B&  3086.796&10:14:08.40&$-01$:35:49.1& 19.52    &  0.90    &$  223.6\pm   4.9$&$ 0.31\pm  0.03$&$ 0.35\pm  0.07$&0.998&$  226.1\pm   1.7$&$ 0.31\pm  0.05$\\
&  15B&  3417.761&&&&&$  226.4\pm   1.8$&$ 0.26\pm  0.03$&$ 0.28\pm  0.06$\\
Sex-0010&   8B&  3086.796&10:14:17.31&$-01$:35:55.0&\nodata   &\nodata   &$  -39.6\pm   2.6$&$ 0.35\pm  0.02$&$ 0.77\pm  0.04$&0.000&$  -39.8\pm   1.6$&$ 0.81\pm  0.02$\\
&  15B&  3417.761&&&&&$  -40.0\pm   2.1$&$ 0.37\pm  0.02$&$ 0.84\pm  0.03$\\
Sex-0011&   8B&  3086.796&10:14:15.20&$-01$:36:16.9& 19.77    &  1.02    &$   72.8\pm   3.0$&$ 0.31\pm  0.04$&$ 0.93\pm  0.06$&0.000&$   64.6\pm   0.8$&$ 0.84\pm  0.04$\\
&  15B&  3417.761&&&&&$   64.0\pm   0.8$&$ 0.37\pm  0.04$&$ 0.73\pm  0.07$\\
Sex-0012&   8B&  3086.796&10:14:21.75&$-01$:31:45.0& 18.12    &  0.82    &$  317.0\pm   1.5$&$ 0.33\pm  0.01$&$ 0.59\pm  0.02$&0.000\\
Sex-0013&   8B&  3086.796&10:14:16.38&$-01$:44:00.5& 20.36    &  1.20    &$  110.9\pm   3.5$&$ 0.43\pm  0.06$&$ 0.56\pm  0.13$&0.000\\
Sex-0014&   8B&  3086.796&10:14:01.64&$-01$:45:32.9& 18.77    &  1.17    &$   16.9\pm   1.8$&$ 0.41\pm  0.03$&$ 0.81\pm  0.04$&0.000&$   16.8\pm   1.4$&$ 0.80\pm  0.03$\\
&  15B&  3417.761&&&&&$   16.5\pm   2.1$&$ 0.36\pm  0.03$&$ 0.79\pm  0.04$\\
Sex-0015&   8B&  3086.796&10:13:55.24&$-01$:33:20.5& 19.80    &  0.95    &$  246.2\pm   4.4$&$ 0.34\pm  0.04$&$ 0.38\pm  0.09$&0.997&$  223.9\pm   1.4$&$ 0.40\pm  0.06$\\
&   7B&  3090.704&&&&&$  233.6\pm   5.1$&$ 0.30\pm  0.10$&$ 0.07\pm  0.25$\\
&  15B&  3417.761&&&&&$  220.2\pm   1.6$&$ 0.34\pm  0.04$&$ 0.44\pm  0.07$\\
...
 \enddata
 \tablenotetext{a}{See electronic edition for complete data table.}
 \label{tab:sex_table}
 \end{deluxetable}

\end{landscape}

\bibliography{ref}

\end{document}